# Observation of anti-levitation of Landau levels in vanishing magnetic fields

W. Pan[1], K.W. Baldwin[2], K.W. West[2], L.N. Pfeiffer[2], and D.C. Tsui[2]

[1] Sandia National Laboratories, Albuquerque, New Mexico 87185, USA
[2] Princeton University, Princeton, New Jersey 08544, USA

We report in this paper an anti-levitation behavior of Landau levels in vanishing magnetic fields in a high quality heterojunction insulated-gated field-effect transistor. We found, in the Landau fan diagram of electron density versus magnetic field, the positions of the magneto-resistance minima at Landau level fillings $\nu$=4, 5, 6 move below the "traditional" Landau level line to lower electron densities. Moreover, the even and odd filling factors show quantitatively different behaviors in anti-levitation, suggesting that the exchange interactions may be important.

Soon after the discovery of the quantum Hall effects (QHEs) [1,2] in two dimensional electron systems (2DES) in high magnetic fields, it was realized that the topological order [3-5] is important in understanding their profound implications. Indeed, unlike all previously known broken symmetry physics, no conventional symmetries are broken in the QHEs. Instead, a new order, the topological order, has to be evoked. It is this topological property that makes the quantization of Hall plateaus insensitive to sample geometries, impurities, densities, etc. In the integer quantum Hall (IQH) effect, the topological property can be quantified by the so-called Chern number [3,4] that can characterize the extended (delocalized) states of a Landau level (LL).

Due to its topological nature, a non-zero Chern number never disappears by itself. This, however, causes a conceptual difficulty in the evolution of a LL as the magnetic field is reduced from a high value to zero. If the extended states stay at the center of the LL, as B → 0, there will be delocalized states below the Fermi surface for a 2DES of finite density. This contradicts the famous scaling theory of Anderson localization [6], which states that at zero temperature all electrons in a two-dimensional system are localized in the absence of magnetic field. Therefore, the question on the fate of the extended states in a LL in vanishing magnetic field naturally arose.

In the mid-'80s, Khmel'nitskii [7] and Laughlin [8] argued that such delocalized states do not disappear. Rather, they float up in energy at small magnetic fields and go to infinite as B → 0. As a result, these delocalized states become inaccessible to the electrons below the Fermi level and the 2DES remains localized. Ten years later, an alternative scenario was provided based on extensive numerical calculations for a tight-binding model (TBM) [9, 10]. There, it was shown that the energy of the delocalized states of Landau levels remains linear with magnetic field until a critical field. Below this critical field, the extended levels disappear. The destruction of the delocalized states in the TBM can be pictured as follows [9,10]: The negative Chern numbers, originally located at the band center, move down with decreasing magnetic field (or increasing disorder), mix with the positive Chern numbers located at the extended levels of lower energy, and eventually annihilate them. Different views on this picture, however, were later offered [11]. More recently, by using the Anderson model on square lattice with on-site random disorder potentials, a surprising anti-levitation behavior was observed from numerical investigations [12].

There, the energy of the extended states of a LL moves below the "traditional" Landau level center as either the disorder strength increases or the magnetic field strength decreases [12]. Similar disorder-driven anti-levitation scenario was already hinted in earlier numerical calculations [13,14].

Experimentally, an early pioneering experiment by Glozman et al [15] apparently confirmed the levitation scenario in their specially designed low mobility GaAs quantum well samples. This claim was further corroborated in Si samples [16]. Later experiments [17-19] on the transitions from the zero-field insulator to IQH states of high LL fillings, however, casted an inconsistent picture with the levitation scenario [20-22]. To our knowledge, there is no experimental report on anti-levitation. Therefore, more than 30 years later, the fate of the delocalized states of Landau levels as B → 0 remains an unsolved problem.

In this paper, we report experimental observation of anti-levitation behavior of Landau levels in vanishing magnetic (B) fields (down to as low as B ~ 58 mT) in a high quality heterojunction insulated-gated field-effect transistor (HIGFET). We observed, in the Landau fan diagram of electron density versus magnetic field, the positions of the magneto-resistance minima at Landau level fillings ν=4, 5, 6 move below the "traditional" Landau level line to lower electron densities. This clearly differs from what was observed in the earlier experiments [15,16] where in the same Landau fan plot the density moved up. Our result strongly supports the anti-levitation behavior [12] predicted recently. Moreover, the even and odd Landau level filling states show quantitatively different behaviors in anti-levitation, suggesting that the exchange interactions, which are important at odd fillings, may play a role.

The specimen used in this study, a HIGFET, is the same as the one in Ref. [23], with a peak electron mobility of ~ $10\times10^6$ cm$^2$/Vs at the electron density n = $1.5\times10^{11}$ cm$^{-2}$. The magneto-resistance $R_{xx}$ was measured by conventional low frequency (~ 11Hz) lock-in technique. Different from conventional quantum Hall measurements, where the magnetic field is swept, in our HIGFET specimen, the gate voltage (or electron density) is swept while the magnetic field is fixed. This measurement setup has a few advantages. For example, with the magnetic field constant, the disorder configuration is fixed and the issue associated with the magnetic field induced disorder potentials is alleviated. Moreover, in this constant magnetic field setup, the separation between Landau levels is constant. Consequently the positions (in energy) of the peaks and valleys of LLs remain unchanged as the electron density is varied. All measurements were taken at the lowest fridge base temperature of 15 mK.

Figure 1 shows $R_{xx}$ as a function of Landau level filling factor $\nu$, which is equivalent to electron density since $\nu=nh/eB$, at a few selected magnetic fields. Curves are shifted vertically for a better view. The oscillatory behavior in $R_{xx}$ is due to the Shubnikov-de Haas oscillations. For high Landau level fillings (e.g., the $\nu=16$ state) the $R_{xx}$ minima stay at the same positions as the magnetic field is varied. For lower LL filings (e.g., the $\nu=4$ state) their minima move to a lower $\nu$ (or n) value as the magnetic field is reduced.

One of the challenges in observing the anti-levitation behavior is to realize a highly uniform 2DES so as to ensure that a density plunging down behavior is not masked by inhomogeneity. In Figure 2, we provide such data confirming a highly uniform 2DES has indeed been achieved in the HIGFET. In Fig. 2a, we plot the electron density versus magnetic fields for a high Landau level filling of $\nu=16$. The straight line represents $n = \nu \times eB/h$. All the data points fall onto the line in the magnetic field range of ~ 58 to 260 mT. To further quantify the tiny density variation in the experimental data, we subtract the theoretically expected value from the data, and plot the difference as a function of magnetic field in Figure 2b. It can be seen that over the experimental B field range, the density difference ($\delta n$) shows no magnetic field dependence and is scattered between $0.1\times10^8$ and $-1\times10^8$ cm$^{-2}$, with a mean value of $<\delta n> \approx -4\times10^7$ cm$^{-2}$. This value represents the uncertainty floor in the electron density, and is much smaller than the lowest density we reached in this study ~ $8\times10^9$ cm$^{-2}$.

We now plot in Fig. 3a the electron density versus magnetic field for three low Landau level fillings at $\nu=4, 5, 6$. A careful examination of the data at $\nu=4, 5, 6$ shows different n versus B dependence, particularly at very low magnetic fields. There, unlike the $\nu=16$ state, the data points clearly deviate from the traditional Landau level linear dependence and move to lower electron densities. Following the same analysis for the $\nu=16$ state, we plot in Fig. 3b the difference between the experimental value and the theoretically expected one for ν=4, 5, 6. For all three fillings, the density plunges down as B is reduced and |δn| increases linearly with decreasing B. At B ~ 0.1T, the difference reaches a value of $4\times10^8$ cm$^{-2}$, 10 times larger than the density uncertainty floor of $4\times10^7$ cm$^{-2}$. In other words, an anti-levitation behavior has been observed for the ν=4, 5, 6 integer quantum Hall states at very low magnetic fields. This observation is consistent with the recent numerical simulations [12-14].

Ideally, we would like to use the peaks in the longitudinal conductivity to track the position of the delocalized states. However, in our high mobility HIGFET, due to the developing features of the fractional quantum Hall effect, for example, between ν=3 and 4 in Fig. 1, determining the positions of the conductivity (or resistivity/resistance) peaks is unreliable. On the other hand, since the magnetic field is constant in our measurements, the positions of the peaks and valleys in the LL spectrum are fixed in energy. Consequently, the valleys, or the positions of magneto-resistance minima, can be viewed as the states (in energy) the most far away from the extended states. In this regard, using the position of a resistance minimum to track the delocalized states of a LL is justified.

The anti-levitation in electron density as B → 0 is surprising. In fact, a floating up in density can be expected due to a large Landau level mixing effect [9,24,25] in the weak field limit, even though there is no levitation in energy [9]. In the HIGFET, the Landau level mixing parameter $\kappa = e^2/\varepsilon l_B/\hbar\omega_c$ is over 10 at B = 58 mT, much larger than the κ achieved in Ref. [15], where $l_B = (\hbar e/B)^{1/2}$ is the magnetic length and $\omega_c$ = eB/m the cyclotron frequency. All other parameters have their normal meanings. In this regard, the observation of a plunging down in density is truly remarkable. This further attests the highest quality of the HIGFET in which an extremely weak field regime can be achieved to reveal the anti-levitation behavior.

It is interesting to notice that the anti-levitation behavior is quantitatively different between the odd and even Landau level fillings. For example, for the ν=5 state, the initial magnetic field at which the anti-levitation behavior starts to develop apparently is higher than that at ν=4 and 6. At the same time, its density plunging rate is lower. We believe that this difference between the even and odd filling factors is due to a different origin in their energy gaps. For the even filling factor states, the gap is a cyclotron gap. For the odd LL filling factors, the gap is due to Zeeman splitting. It is known that the exchange interactions are important for Zeeman splitting enhancement at small odd fillings. Considering these, our data in Figure 3 indicate that the exchange interactions, which have not been considered in the tight-banding numerical calculations, must play a role in the anti-levitation behavior. Finally, we note that the $\nu=4$ and 6 states show more or less the same anti-levitation behavior.

In summary, an anti-levitation behavior of Landau levels in vanishing magnetic fields was observed in a high quality heterojunction insulated-gated field-effect transistor, consistent with recent numerical calculations. Moreover, the even and odd filling factor states show quantitatively different behaviors, suggesting that the exchange interactions may play an important role in anti-levitation.

We thank X.R. Wang and K. Yang for very helpful discussions. This work was supported by the U.S. Department of Energy, Office of Science, Basic Energy Sciences, Materials Sciences and Engineering Division. Sandia National Laboratories is a multi-program laboratory managed and operated by Sandia Corporation, a wholly owned subsidiary of Lockheed Martin Corporation, for the U.S. Department of Energy's National Nuclear Security Administration under contract DE-AC04-94AL85000. Sample growth at Princeton was funded by the Gordon and Betty Moore Foundation through the EPiQS initiative GBMF4420, and by the National Science Foundation MRSEC Grant DMR-1420541.

**Figures and figure captions**

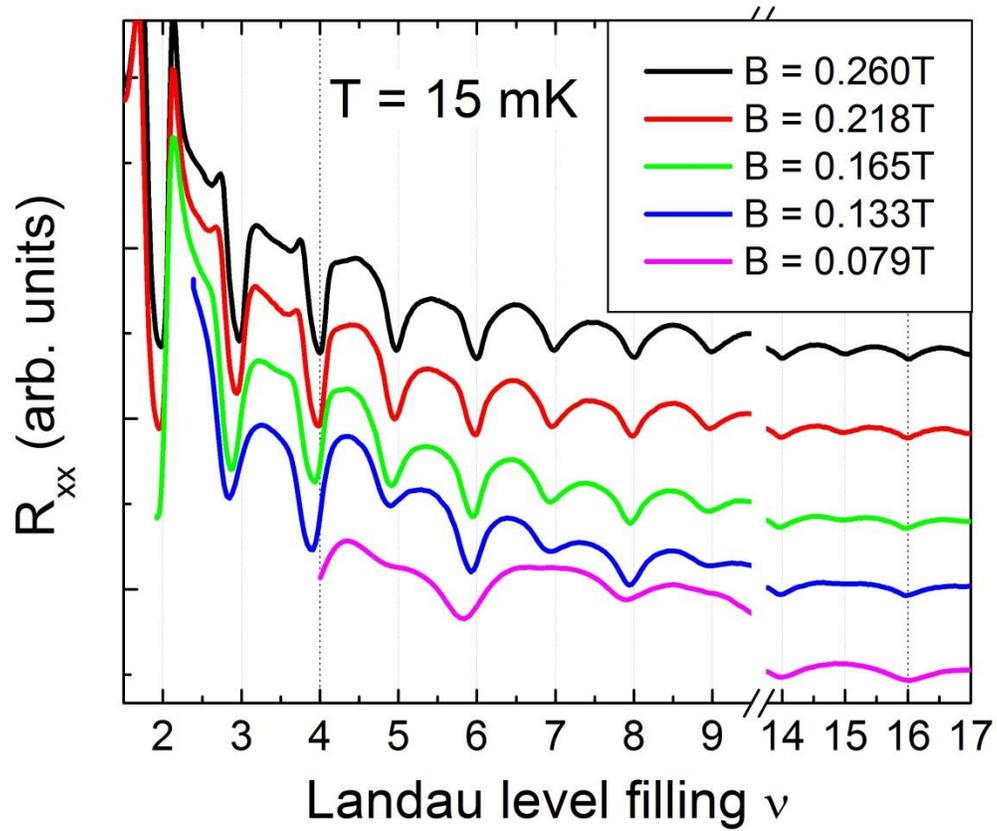

Fig. 1: $R_{xx}$ versus $\nu$ at different magnetic fields. Traces are shifted vertically for clarity.

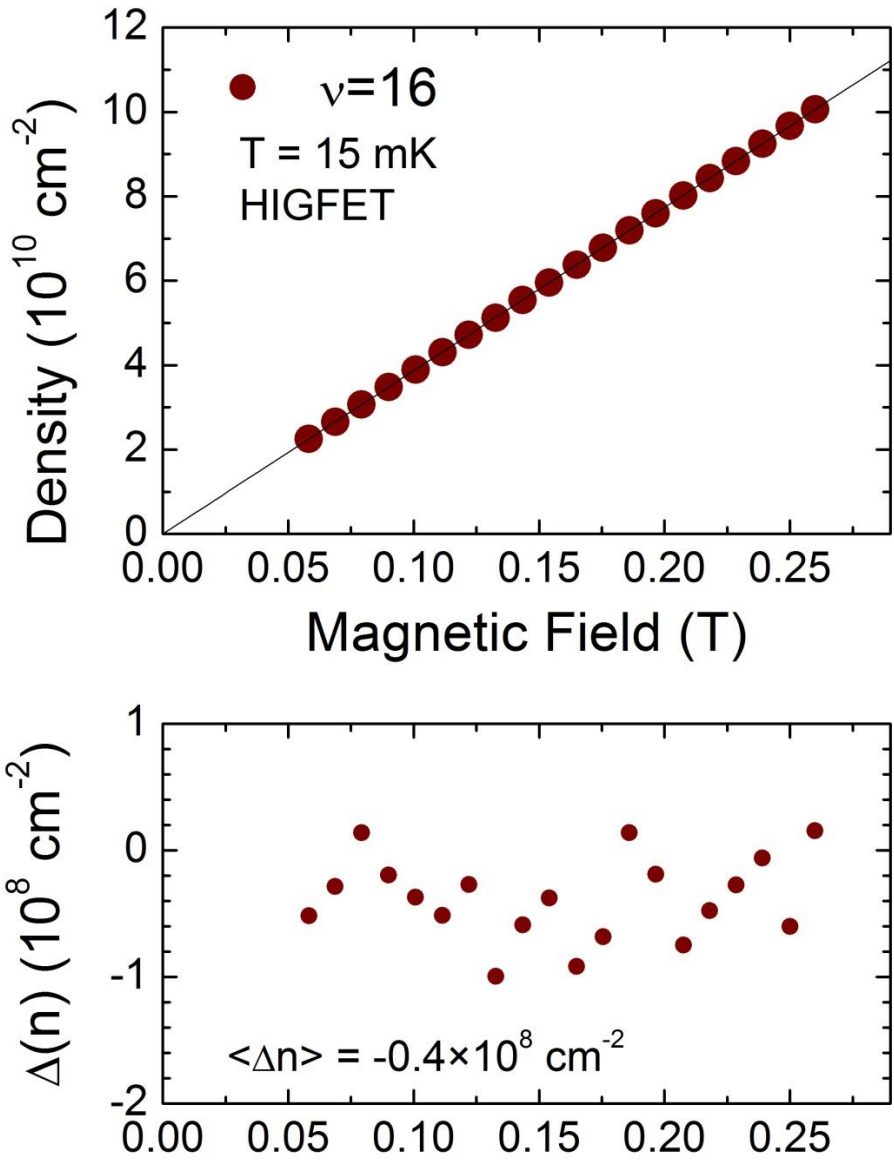

Fig. 2: (a) Landau fan diagram of electron density versus magnetic field for the ν=16 state. (b) Density difference between the experimentally measured values and the theoretically expected values.

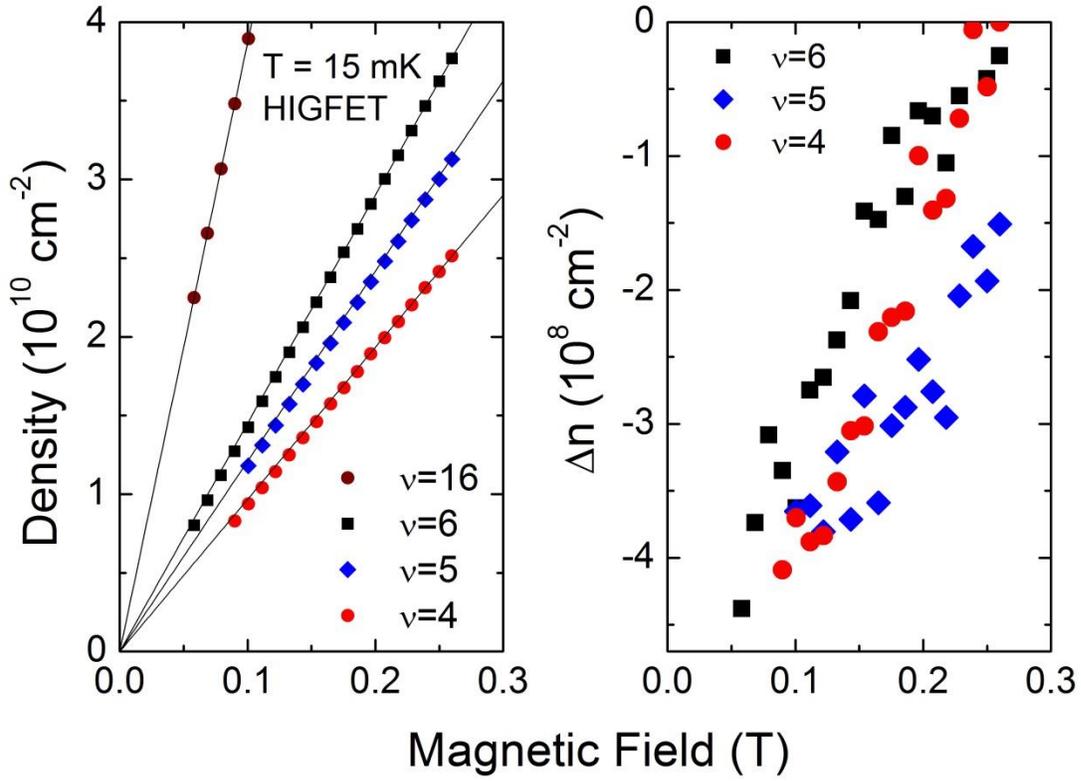

Fig. 3: (a) Landau fan diagram for the $\nu=4, 5, 6$ states. Data for the $\nu=16$ state are also included for comparison. (b) Density difference between the experimentally measured values and the theoretically expected values for the $\nu=4, 5, 6$ states.